\documentclass[%
 aip,
 jmp,%
 amsmath,amssymb,
 reprint,%
]{revtex4-1}

\usepackage{graphicx}
\usepackage{dcolumn}
\usepackage{bm}

\usepackage{verbatim}
\newcommand{\code}[1]{\texttt{#1}}

\newcommand{\bbx}{BeatBox}
\newcommand{\bbg}{\code{.bbg}} 

\newcommand\pd{\partial}		

\newcommand{\df}[2]{\frac{\pd{#1}}{\pd{#2}}} 

\def\eqreftwo(#1,#2){(\ref{eq:#1},\ref{eq:#2})}

\newcommand{\kron}{\delta}              

\newcommand{\mx}[1]{\mathbf{#1}}        

\newcommand{\Real}{\mathbb{R}}          

\newcommand{\+}[2]{\def#1{{#2}}}
\newcommand{\1}[2]{\def#1##1{{#2}}}

\+{\num}{^{\sharp}}           
\+{\Boundary}{\Gamma}         
\+{\C}{C}                     
\+{\CM}{C_m}                  
\+{\chf}{\psi}                
\+{\coeff}{\nu}               
\+{\D}{D}                     
\+{\Dten}{\hat{D}}            
\+{\Deten}{\hat{D}^e}         
\+{\Diten}{\hat{D}^i}         
\+{\Ppar}{P_{\parallel}}      
\+{\Port}{P_{\perp}}          
\+{\Depar}{D^e_{\parallel}}   
\+{\Deort}{D^e_{\perp}}       
\+{\Dipar}{D^i_{\parallel}}   
\+{\Diort}{D^i_{\perp}}       
\+{\Deeff}{D^e_{*}}           
\+{\Dieff}{D^i_{*}}           
\+{\Deff}{D_{*}}              
\+{\Domain}{\mathcal{D}}      
\+{\di}{\Delta_i}             
\+{\dj}{\Delta_j}             
\+{\dk}{\Delta_k}             
\+{\dt}{k}                    
\+{\dx}{h}                    
\1{\E}{\times10^{#1}}         
\+{\err}{\varepsilon}         
\+{\f}{\mx{f}}                
\+{\fib}{f}                   
\+{\fibvec}{\vec{\fib}}       
\+{\fnb}{\beta}               
\+{\fng}{\gamma}              
\+{\fne}{\epsilon}            
\+{\g}{\mx{g}}                
\+{\Ie}{I_{\mathrm{ext}}}     
\+{\Ieff}{I_{\mathrm{eff}}}   
\+{\Iion}{I_{\mathrm{ion}}}   
\+{\Iu}{I_u}                  
\+{\Iv}{I_v}                  
\+{\i}{i}                     
\+{\j}{j}                     
\+{\k}{k}                     
\+{\L}{\mathcal{L}}           
\+{\Len}{L}                   
\+{\Linf}{L^\infty}           
\+{\Ltwo}{L^2}                
\+{\n}{n}                     
\+{\normv}{\vec{n}}           
\+{\norm}{n}                  
\+{\r}{{\vec r}}              
\+{\sig}{\sigma}              
\+{\sigi}{\hat{\sig}_{\mathrm{i}}} 
\+{\sige}{\hat{\sig}_{\mathrm{e}}} 
\+{\sigeff}{\hat{\sig}_{\mathrm{eff}}} 
\+{\svr}{\chi}                
\+{\T}{T}                     
\+{\t}{t}                     
\+{\u}{u}                     
\+{\V}{V}                     
\+{\Va}{V^*}                  
\+{\v}{v}                     
\+{\weight}{W}                
\+{\p}{\mx{p}}                
\+{\phie}{\Phi_e}             
\+{\phii}{\Phi_i}             
\+{\phiia}{\Phi^*_i}          
\+{\pin}{_{\textrm{pin}}} 
\+{\q}{\mx{q}}                
\+{\x}{x}                     
\+{\xc}{x_*}                
\+{\y}{y}                     
\+{\yc}{y_*}                
\+{\z}{z}                     
\+{\A}{A}                     
\+{\B}{B}                     
\+{\mS}{\mx{S}}               
\+{\mD}{\mx{\Lambda}}         
\+{\mT}{\mx{T}}               
\usepackage{color}
\newcommand{\chg}[2][0]{#2}
\newcommand{\del}[1]{}
\renewcommand{\u}{\mx{u}}               
\renewcommand{\D}{\mx{D}}                 
\newcommand{\Q}{\mx{Q}}                 
\newcommand{\paralp}{\alpha}            
\newcommand{\parbet}{\beta}             
\newcommand{\pargam}{\gamma}            

\begin{document}

\preprint{AIP/123-QED}

\title[]{Cardiac re-entry dynamics \&
  self-termination in DT-MRI based model of
  Human Foetal Heart}%
\thanks{As submitted to Chaos: An Interdisciplinary Journal of
  Nonlinear Science, Focus Issue on the topic of Complex Cardiac Dynamics.}

\author{I.V.Biktasheva}
\altaffiliation[Also at ]{CEMPS, University of Exeter, Exeter EX4 4QF,
  UK. 
\\Author to whom correspondence should be addressed}%
 \email{ ivb@liv.ac.uk}
 \affiliation{Department of Computer Science, 
   University of Liverpool, Liverpool L69 3BX, UK}%
\author{R.A.Anderson}
 \affiliation{MRC Centre for Reproductive Health,
   University of Edinburgh, Edinburgh EH16 4T3, UK}%
\author{A.V.Holden}
 \affiliation{School of Biomedical Sciences, University of Leeds,
   Leeds LS2 9JT, UK}%
\author{E.Pervolaraki}
 \affiliation{School of Biomedical Sciences, University of Leeds,
   Leeds LS2 9JT, UK}%
\author{F.C.Wen}
 \affiliation{Department of Computer Science, 
   University of Liverpool, Liverpool L69 3BX, UK}%

\date{\today}

\begin{abstract}
The effect of heart geometry and anisotropy on cardiac re-entry
dynamics and self-termination is studied here in anatomically
realistic computer simulations of human foetal heart.
20 weeks of gestational age human foetal heart isotropic and anisotropic anatomy models
from diffusion tensor MRI data sets are used in the computer
simulations. The fiber orientation 
angles of the heart were obtained from the DT-MRI primary eigenvalues. In a spatially
homogeneous electrophysiological mono domain model with the DT-MRI
based heart geometries, we initiate simplified Fitz-Hugh-Nagumo kinetics cardiac re-entry at a
prescribed location in a 2D slice, and in the full 3D anatomy model. In a slice of the heart, the MRI based fiber anisotropy changes the re-entry dynamics
from pinned to anatomical re-entry. In the full 3D MRI based model,
the foetal heart fiber
anisotropy changes the re-entry dynamics
from a persistent re-entry to the re-entry self-termination. Time of
re-entry self-termination depends on the re-entry initial
position. Anisotropy of the heart speeds up re-entry self-termination.  
\end{abstract}

\maketitle

\begin{quotation}
The effect of the heart anisotropy and anatomy on cardiac re-entry
dynamics, although difficult to demonstrate in experiment, is well appreciated~\cite{Bishop-etal-2010,Bishop-etal-2011,Bishop-Plank-2012,Fukumoto-etal-2016},
and has been studied in simplified mathematical and computer
models~\cite{Fenton-Karma-1998,Pertsov-etal-PRL2000,Wellner-etal-PNAS2002,
RodriguezEasonTrayanova-2006,Dierckx-etal-PRE2013}. The 
\bbx~\cite{bbx-2017-PONE} High Performance Computing (HPC) 
cardiac electrophysiology computer simulation environment allows
direct incorporation of the high resolution
DT-MRI heart anatomy data sets into the biophysically and anatomically realistic
computer simulations.   In the \bbx ~\emph{in-silico} model, the anisotropy of the
tissue is 
switched ``on'' and ``off'' to allow for comparison between the anatomically realistic
isotropic and anisotropic conduction, in order to see the specific
pure anatomy
effects, as well as the interplay
between the anisotropy and anatomy of an individual heart.
In this paper, we present the DT-MRI based anatomy and myofiber structure  
realistic computer simulation study of cardiac re-entry dynamics in
the \emph{in-silico} model of the human foetal heart~\cite{Pervolaraki-etal-2013}. We demonstrate that,
in a 2D slice of the heart, the realistic fiber anisotropy of the
tissue changes cardiac re-entry dynamics
from pinned into fast anatomical re-entry. In the full 3D DT-MRI based model, depending on
the initial location of the re-entry, the isotropic geometry of the heart might
sustain a perpetual re-entry even with a positive filament tension; while the same positive
filament tension re-entry initiated at the same
location of the foetal heart with the realistic fiber
anisotropy self-terminates within seconds. Generally, time of
re-entry self-termination depends on the re-entry initial
position, while the role of the heart anisotropy is to speed up the re-entry self-termination.  
\end{quotation}

\section{\label{Intro}Introduction}

Since the hypothesis over a century ago that cardiac re-entry
underlies cardiac arrhythmias~\cite{Mines-1913, Garey-1914} , and the  
much later confirmation of the hypothesis in cardiac tissue
experiment~\cite{Allessie-etal-1973, Pertsov-etal-1993}, 
the re-entry (\emph{aka} spiral wave in 2D, cardiac excitation vortex in
3D), its origin and its role in sustained arrhythmias and fibrillation,
as well as a possibility of its 
effective control and defibrillation, have
been an object of extensive theoretical study and
modelling~\cite{Wiener-Rosenblueth-1946,Balakhovsky-1965,Krinsky-1968,Panfilov-etal-1984,Davydov-etal-1988,Keener-1988,Ermakova-etal-1989,Biktashev-Holden-1994,awt,Fenton-Karma-1998,Pertsov-etal-PRL2000,Wellner-etal-PNAS2002,swd,orbit,Biktashev-etal-2011-PONE,Biktasheva-etal-2015-PRL}. 
From experiment, it is an established point of view that cardiac arrhythmias are due to a
complex combination of
electrophysiological~\cite{BoschNattel-CardiovascularResearch2002,Workman-etal-HeartRythm2008,Kushiyama-etal-2016}, structural~\cite{Pellman-etal-2010,Eckstein-etal-2011,Takemoto-etal-2012,Eckstein-etal-2013},
and anatomical~\cite{MacEdo-etal-2010,Anselmino-etal-2011} factors
which sustain cardiac re-entry~\cite{GrayPertsovJalife-1996-Circ,Wu-etal-1998-CR,Nattel-Nature2002,Yamazaki-etal-2012-CVR}.

The specific effect of the heart anisotropy and anatomy on cardiac re-entry
dynamics is well appreciated~\cite{Bishop-etal-2010,Bishop-etal-2011,Bishop-Plank-2012,Fukumoto-etal-2016},
and has been studied in simplified mathematical and computer
models~\cite{Fenton-Karma-1998,Pertsov-etal-PRL2000,Wellner-etal-PNAS2002,
RodriguezEasonTrayanova-2006,Dierckx-etal-PRE2013}.
\chg[mechanistic]{
The anisotropic discontinuities in the heart muscle have been commonly seen as
a substrate for rise of cardiac re-entry due to the abrupt change in
conduction velocity and wavefront curvature\cite{Fenton-Karma-1998, 
  Spach-CircRes-2001, Smaill-etal-2004}. On the other hand, \chg[incidental]{extensive mapping of cardiac
myocyte orientation in mammalian hearts has shown that the transmural fiber
arrangement, including the range of transmural change in fiber angle in ventricular wall, was consistent within a
species, and varied between  
species\cite[p.~173]{Hunter-etal-CompBiolOfHeart}. }
So that changes in anisotropy seen in healthy hearts can facilitate initiation of arryhthmias.
}

The recent advance in 
DT-MRI technology and High Performance Computing (HPC) allows the
obtained DT-MRI data sets  with the detailed heart anatomy and myofiber
structure to be directly incorporated into the anatomically realistic
computer simulations~\cite{bbx-2017-PONE}, so that the anisotropy of the
tissue in the \emph{in-silico} model can be
switched on and off to allow for comparison between the anatomically realistic
isotropic and anisotropic conduction in order to see specific anatomy
effects as well as the interplay
between the anisotropy and anatomy of an individual heart.

In this paper, we present the raw DT-MRI based anatomically and myofiber structure  
realistic computer simulation study of cardiac re-entry dynamics in
the \emph{in-silico} model of human foetal heart. 
\chg[segmentation]{\chg[segmentation-2]{\chg[epi/endo]{
The raw DT-MRI image data~\cite{Pervolaraki-etal-2013} was segmented
into the tissue/non-tissue pixels 
based on the MRI luminosity threshold, followed by the calculation of
the fiber angles at each voxel from the diffusion-weighted DT-MRI images.}
This very basic segmentation}} might be seen as a limitation of
  the study from the cardiac physiology point of view. However, the
  purpose of our study is not to provide results of immediate
  physiological or clinical relevance: for these we currently simply
  have not enough data. Rather,  from the non-linear science point
of view our rationale is to use the raw DT-MRI data ``as is'' as an
example of an unaltered  nature provided medium to study a re-entry
dynamics. Although the DT-MRI yields three eigenvalues,
the second and the third are often harder to
distinguish, so we used only the primary eigenvector to
define the local fibre orientations in the simulation study. 
The focus of the paper is to demonstrate the effect of a real
mammalian heart anatomy and anisotropy on a re-entry dynamics. The
available MRI data of a foetal heart provide an excellent oportunity
for such study. The objectives for the use of foetal heart MRI data
are: whether the anatomical settings of the although foetal but a real heart might support
a positive filament tension re-entry, and what would it be the role of
a real heart anisotropy in that case. So, here we demonstrate that the real
heart anisotropy enhances re-entry self-termination. 

We demonstrate that,
in a 2D slice of the heart, the realistic fiber anisotropy might change the re-entry dynamics
from pinned to anatomical re-entry. 

In the full 3D DT-MRI based model, depending on
the location of the re-entry initiation, the isotropic geometry of the heart might
sustain perpetual re-entry even with a positive filament tension
kinetics. While the same positive
filament tension re-entry initiated at the same
location of the foetal heart with the realistic fiber
anisotropy self-terminates within seconds. Time of
re-entry self-termination depends on the re-entry initial
position. Anisotropy of the real heart speeds up re-entry self-termination,
and in this sense has a rather
  anti-arrhythmogenic effect.  \chg[mechanistic]{The geometry and 
    anisotropy of the heart together 
    ensure the fastest self-termination of cardiac re-entry}. 

\chg[novelty]{The novel significance of our findings is
that
we demonstarte that the real life heart anisotropy might
have a  
rather anti-arrhythmic function as it facilitates fast self-termination of cardiac
re-entry. }

\section{\label{Methods}Methods}

\subsection{\label{DT-MRI}DT-MRI based anatomy model}
\begin{figure}[tb] \centering
\includegraphics{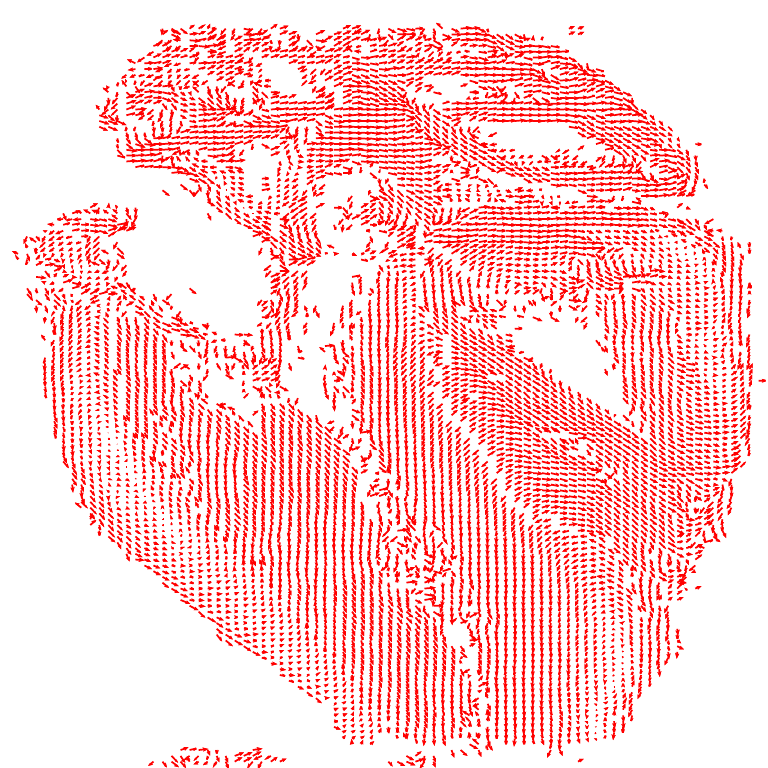}%
\caption{\label{HF_slice_x_63} \chg[epi/endo]{{\bf The 143 DGA human foetal heart~\cite{Pervolaraki-etal-2013}}.
  \bbx~\cite{bbx-2017-PONE} geometry \code{.bbg} format
  visualisation\chg[color-coding]{: shown here are projections of the unit vectors of the local fibre orientation onto the  cross-section plane.
Laminar fibres are well formed except near the surfaces of the outer walls,  where chaotic structure from earlier developmental stages is still present. 
This is seen even better in the colour-coded  Figure 4 in Pervolaraki et al~\cite[p.~5]{Pervolaraki-etal-2013}. 
}}}
\end{figure}

The DT-MRI data sets of the $128\times128\times128$ voxels size, with voxel resolution of
$\sim100\mu m$, of ethically obtained 143 days of gestational age (DGA) human foetal
heart~\cite{Pervolaraki-etal-2013}, were converted into the
\bbx~\cite{bbx-2017-PONE} regular Cartesian mesh \code{.bbg} geometry
format, containing the DT-MRI cartesian coordinates of the heart
tissue points together with the
corresponding components of the diffusion tensor primary
eigenvectors~\cite{bbx-2017-PONE}. The \bbg\ file is an ASCII text
file,  each line in
which describes a point in a regular mesh in the following format: %
\[
 \code{ x,y,z,status,fibre\_x,fibre\_y,fibre\_z}
\]
Here \code{x, y, z} are integer Cartesian coordinates of a DT-MRI voxel,
\code{status} is a flag with a nonzero-value for a tissue point, and \code{fibre\_x, fibre\_y, fibre\_z} are $x$-,
$y$- and $z$-components of the fibre orientation vector at that
point. To reduce the size of the \bbg\ files, only the tissue points,
that is points with nonzero \code{status} need to be specified,
because the \bbx\ solver
will ignore the void points with zero status in any case. Although
the original DT-MRI images data sets had $128\times128\times128$ voxels size,
the actual dimensions of the foetal heart minimum bounding box were
$67\times91\times128$, with $181070$ tissue points.

\chg[segmentation]{\chg[segmentation-2]{\chg[epi/endo]{
The raw DT-MRI anatomy data~\cite{Pervolaraki-etal-2013} were
segmented into the ``tissue''/``not
tissue'' pixels discretion
based on the MRI luminosity threshold, with the cartesian fiber angles at each voxel
obtained from the diffusion-weighted DT-MRI images. Only this basic
segmentaion of the raw DT-MRI anatomy
data~\cite{Pervolaraki-etal-2013} was taken into account in the
computer simulation of cardiac re-entry dynamics, so we shall refer to
it as the raw DT-MRI based anatomy model.}}}

\chg[2Dfibers]{In the 2D model, the fibres vectors were projected into the plane, in order to
construct the 2D diffusivity tensor. }

\chg[incidental]{\chg[epi/endo]{\chg[color-coding]{ 
Fig.~\ref{HF_slice_x_63} %
 shows the cross section of the 143 days of gestational age (DGA) foetal
 heart with already formed intramural laminar
 structure and more irregular epicardial, endocardial, and septal fibers, see Figure 4 of Pervolaraki et al~\cite[p.~5]{Pervolaraki-etal-2013}  for the color-encoded fractional anisotropy (FA) and all the three components of
      the fiber angles in human foetal hearts. The DT-MRI based foetal
    heart model}} offered a unique opportunity to see if the 20 weeks of gestation age intramural heart structure was capable to support
cardiac re-entry, as it would not be possible for the re-entry to pin
to the endocardial fine features which were yet to be
developed later, such as \emph{e.g.} the pinning
to pectinate muscles junction with crystae terminalis reported in adult
human atria~\cite{Wu-etal-1998-CR,Yamazaki-etal-2012-CVR,Kharche-etal-2015-BMRI}. }

\subsection{\label{RD} Cardiac Tissue Model}

To investigate the effects of anatomy on cardiac re-entry dynamics we used
\textit{monodomain} tissue model with non-flux boundary conditions 
\begin{align} &
  \df{\u}{\t} = \f (\u)
  + \nabla\cdot\hat\D\nabla \u ,  
			    \label{bc} \\ &
  \qquad \qquad {\normv \cdot\hat\D\nabla \u}\bigg|_G = 0,  
			    \nonumber 
\end{align}
where  $\u(\r,t)={(u, v)}^T$, $\r$ is the position vector, $\f(\r,t)={(f, g)}^T$	
is the Fitz-Hugh-Nagumo~\cite{Winfree-1991} kinetics column-vector 
\begin{align}
  f(u,v) &= \paralp^{-1}(u-u^3/3-v),                         \nonumber\\
  g(u,v) &= \paralp \, (  u + \parbet - \pargam v ), \label{FHN}
\end{align}
 with the parameter values $\paralp=0.3$, $\parbet=\chg[]{0.71}$,
 $\pargam=0.5$, which in an infinite excitable medium support a rigidly rotating vortex with positive filament
 tension~\cite{ft}.
The simplified FHN model was intentionally chosen for
  this study in order to fully eliminate
 the possible effects of a realistic cell excitation kinetics, such as \emph{e.g.}
 meander~\cite{Winfree-1991}, alternans\cite{Karma_Chaos1994}, negative filament tension\cite{ft},
 etc., and in order to enhance and highlight the pure effects of the heart
 anatomy and anisotropy on the cardiac re-entry outcome.
$\hat\D=\Q\hat P$, where $\Q=\mathrm{diag}(1,0)=\begin{bmatrix}1&0\\0&0\end{bmatrix}$ is the matrix of
the relative diffusion coefficients for $u$ and $v$ components, and $\hat
P=[P_{\j\k}]\in\Real^{3\times 3}$ is the $u$ component diffusion tensor,
which has
only two different eigenvalues: the bigger, simple eigenvalue $\Ppar$
corresponding to the direction along the tissue fibers, and the
smaller, double eigenvalue $\Port$, corresponding to the directions
across the fibres, so that
\begin{equation}
  P_{\j\k} = \Port\kron_{\j\k} +
  \left(\Ppar-\Port\right)\fib_\j\fib_\k,
                                        \label{sigma}
\end{equation}
where $\fibvec=\left(\fib_\k\right)$ is the unit vector of the fiber
direction; $\normv$ is the vector normal to the tissue boundary
$G$. In the isotropic
simulation, $\Ppar$ and $\Port$ values were fixed at
$\Ppar=\Port=1$ (corresponding 1D conduction velocity \chg[]{1.89}). In the anisotropic simulations, $\Ppar$ and $\Port$
values were fixed at $\Ppar=2$, $\Port=0.5$ (corresponding
  conduction velocities \chg[]{2.68} and \chg[]{1.34} respectively). All the
  conduction velocities have been computed for the period waves with
  the frequency of the free spiral wave in the model,
  i.e. \chg[]{11.36}. \chg[CV]{\chg[average]{With the isotropic
diffusivity ($\Ppar=\Port=1$) equal to the geometric mean between
the faster and the slower anisotropic diffusivities ($\Ppar=2, \Port=1/2$), the isotropic conduction
velocity 1.89 was almost exactly the same as the geometric mean 1.89 of the
faster and slower (2.68 and 1.34 respectively) anisotropic
conduction velocities, chosen in order to minimize the maximal relative difference between the isotropic and anisotropic propagation
speeds.}}

All the computer simulations presented here were done using the \bbx~\cite{bbx-2017-PONE} 
software package with the explicit time-step Euler
scheme, on the Cartesian regular grid with space step discretization $\Delta x=0.1$, time step
discretisation $\Delta t=0.001$; 5-point stencil for isotropic, and
9-point stencil for anisotropic Laplacian approximation in 2D
simulations; 7-point stencil for isotropic, and 27-point stencil for anisotropic Laplacian approximation in 3D
simulations. The re-entry was initiated by the phase distribution
method~\cite{chaos}: in the 2D simulations, at a prescribed location of the
cross section of the DT-MRI based
anatomical model; in the 3D simulations, at a prescribed location of the full DT-MRI based whole heart
anatomical model.  


\section{\label{RESULTS}Results}

\subsection{\label{2D} 2D MRI-based ``slice'' simulations}
\begin{figure*}[tb]
\includegraphics{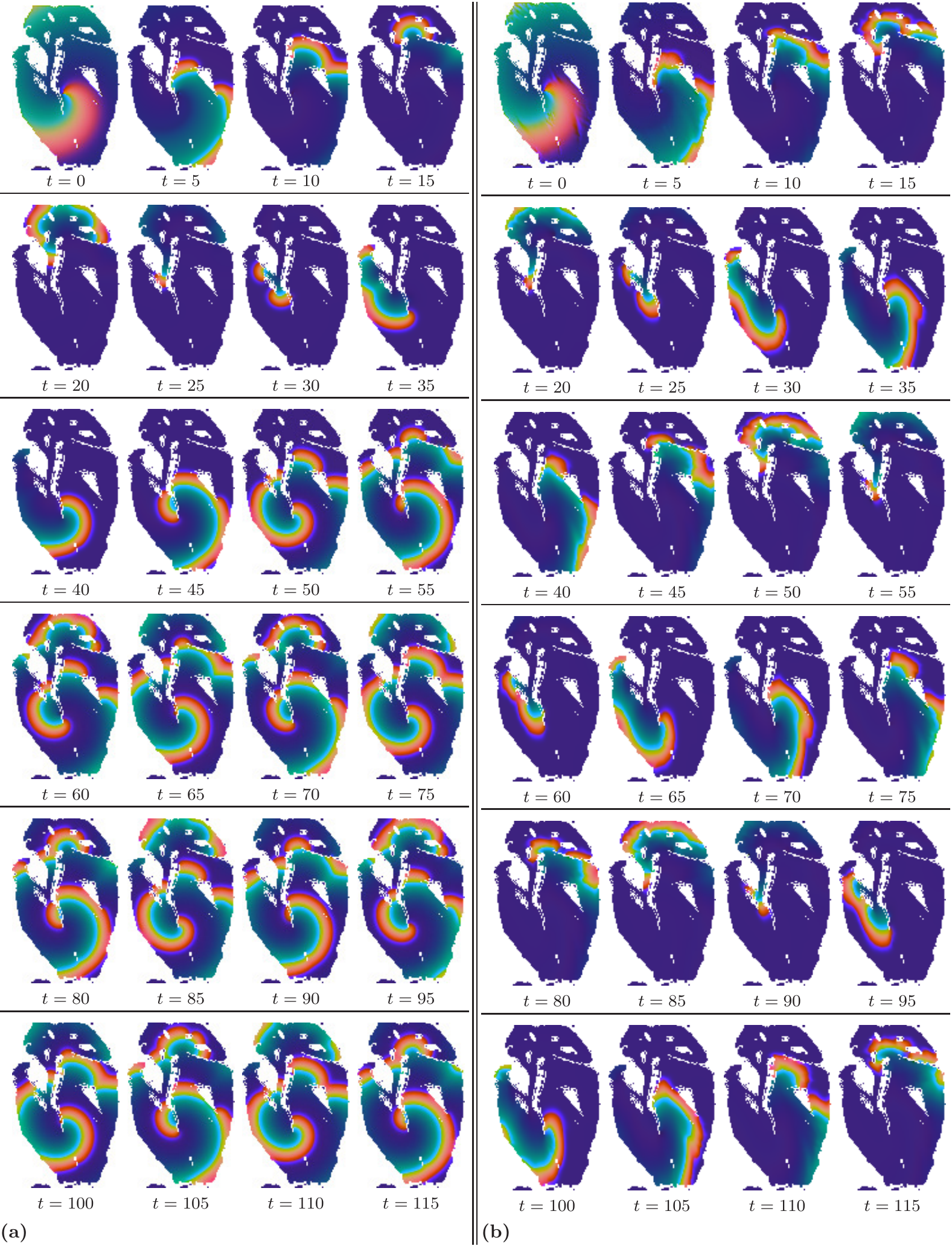}%
\caption{\label{slice_fig} {\bf Anisotropy effect in the 2D slice
    simulations}, time shown \chg[units]{under each
  panel in time units of Eqs.~(\ref{bc})-(\ref{FHN})}. {\bf a)} \emph{Isotropic conduction}: after the transient first rotation around the septum cuneiform
 opening, the slow excitation re-entry pins to the 
 sharp low end of the opening in the foetal heart. (Multimedia view)
 Fig2a.mpg. {\bf b)}
  \emph{Anisotropic conduction}: 
after the fast
 transient first rotation around the septum cuneiform
 opening, the anisotropy of the foetal heart turns
 the initial spiral wave into the fast anatomical re-entry around 
 the septum cuneiform
 opening. (Multimedia view) Fig2b.mpg.}
\end{figure*}

In the 2D simulations, Fig.~\ref{slice_fig}, a counter-clockwise re-entry was initiated by the phase distribution
method~\cite{chaos},  with the initial center of rotation placed at the
prescribed location $x_0=40, y_0=60$ in the 2D cross section of the DT-MRI based
anatomical model shown in Fig.~\ref{HF_slice_x_63}. 

In the Fig.~\ref{slice_fig}(a-b),  %
it can be seen
that in both isotropic and anisotropic 2D simulations, at $t=0$, there
was identical location of  the initial re-entry
rotation center: roughly in the middle of
the slice, in the vicinity of the septum cuneiform opening.  

Fig.~\ref{slice_fig}(a) shows the isotropic dynamics of the re-entry,
that is 
with the fiber orientation data ``turned OFF'', so that only the
geometry of the isotropic homogeneous slice affects the dynamics of the re-entry. While it
is known that in an infinite medium the chosen FHN kinetics parameter values $\paralp=0.3$, $\parbet=\chg[]{0.71}$,
 $\pargam=0.5$ produce rigidly rotating spiral~\cite{Winfree-1991}, the anatomically
 realistic boundaries of the foetal heart cause the drift of the
 re-entry. The re-entry does not terminate because of the resonant
 reflection from the inexcitable boundaries~\cite{Biktashev-Holden-1994}, but
 after the transient first rotation around the septum cuneiform
 opening, the tip of the re-entry firmly pins to the sharp lower end of the cuneiform
 opening, see Fig.~\ref{slice_fig}(a).

Fig.~\ref{slice_fig}(b) shows the anisotropic dynamics of the re-entry,
that is 
with the fiber orientation data ``turned ON'', so that both the
anatomically realistic 
geometry and the anisotropy of the otherwise homogeneous slice of the
heart affect the dynamics of the re-entry, causing its drift. In the
anisotropic slice, the re-entry also does not terminate at the inexcitable boundaries, but
 after the faster than in the isotropic case, see the $a.u.$ time
 labels in the Fig.~\ref{slice_fig}(a-b),
 transient first rotation around the septum cuneiform
 opening, the anatomically realistic anisotropy of the medium turns
 the initial spiral wave into the fast anatomical re-entry around 
 the septum cuneiform
 opening, see Fig.~\ref{slice_fig}(b).

\subsection{\label{3D} 3D Whole heart MRI-based simulations}
\begin{figure*}[tb]
\includegraphics{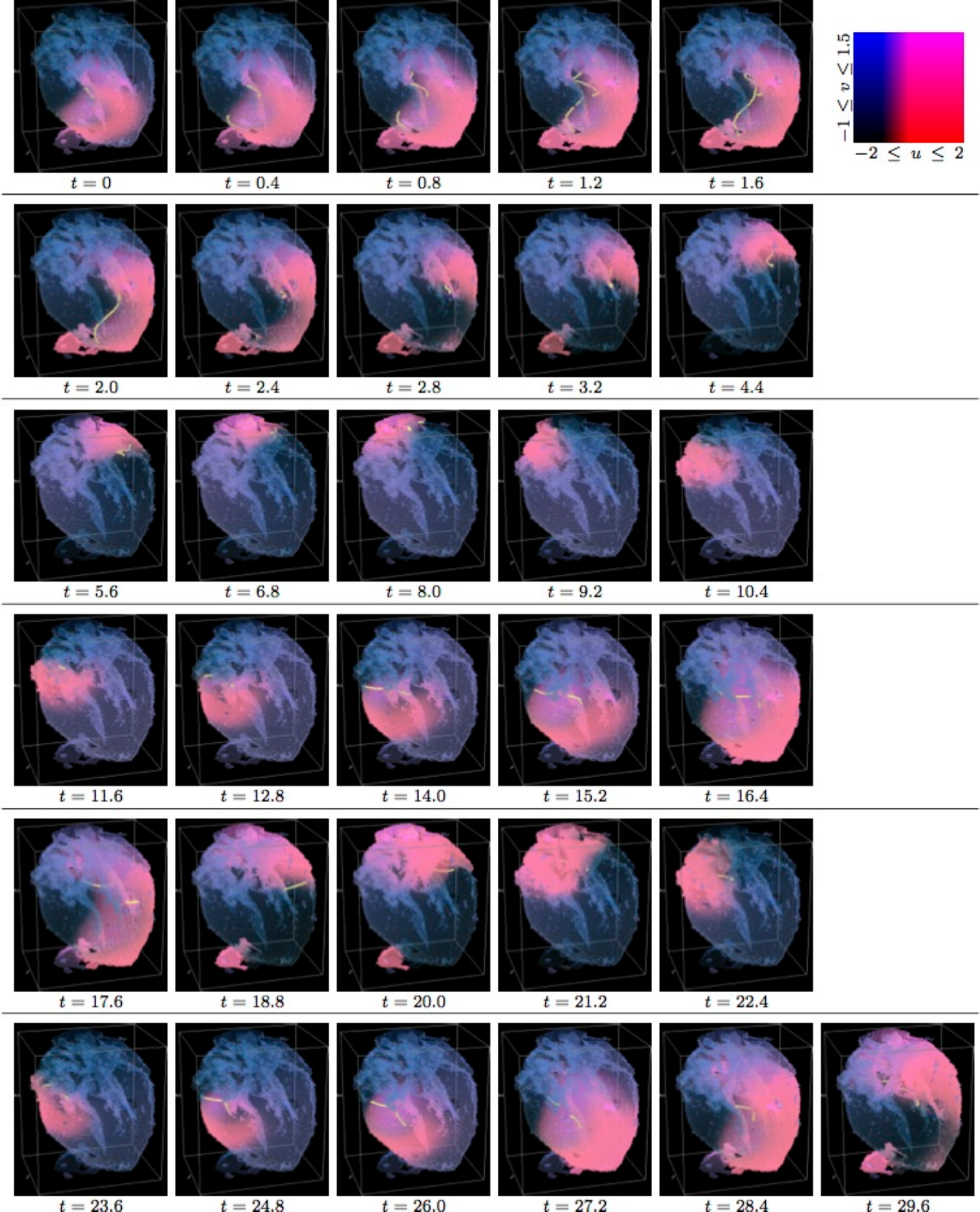}%
\caption{\label{3D_iso} {\bf Isotropic whole heart 
    simulation.} The translucent foetal heart is shown
  in blue, excitation front shown in red (see the color box in FIG.~\ref{3D_iso}), the yellow lines are the
  instant organising filaments of the excitation
  vortices; time shown \chg[units]{under each
  panel in time units of Eqs.~(\ref{bc})-(\ref{FHN})}. After a short transient the organising filament of the initial vortex breaks into the
 two short pieces each of which finds its own synchronous perpetual
 pathway, resulting in the perpetual cardiac re-entry in the foetal
 heart. (Multimedia view) Fig3.mpg. }
\end{figure*}

\begin{figure*}[tb]
\includegraphics{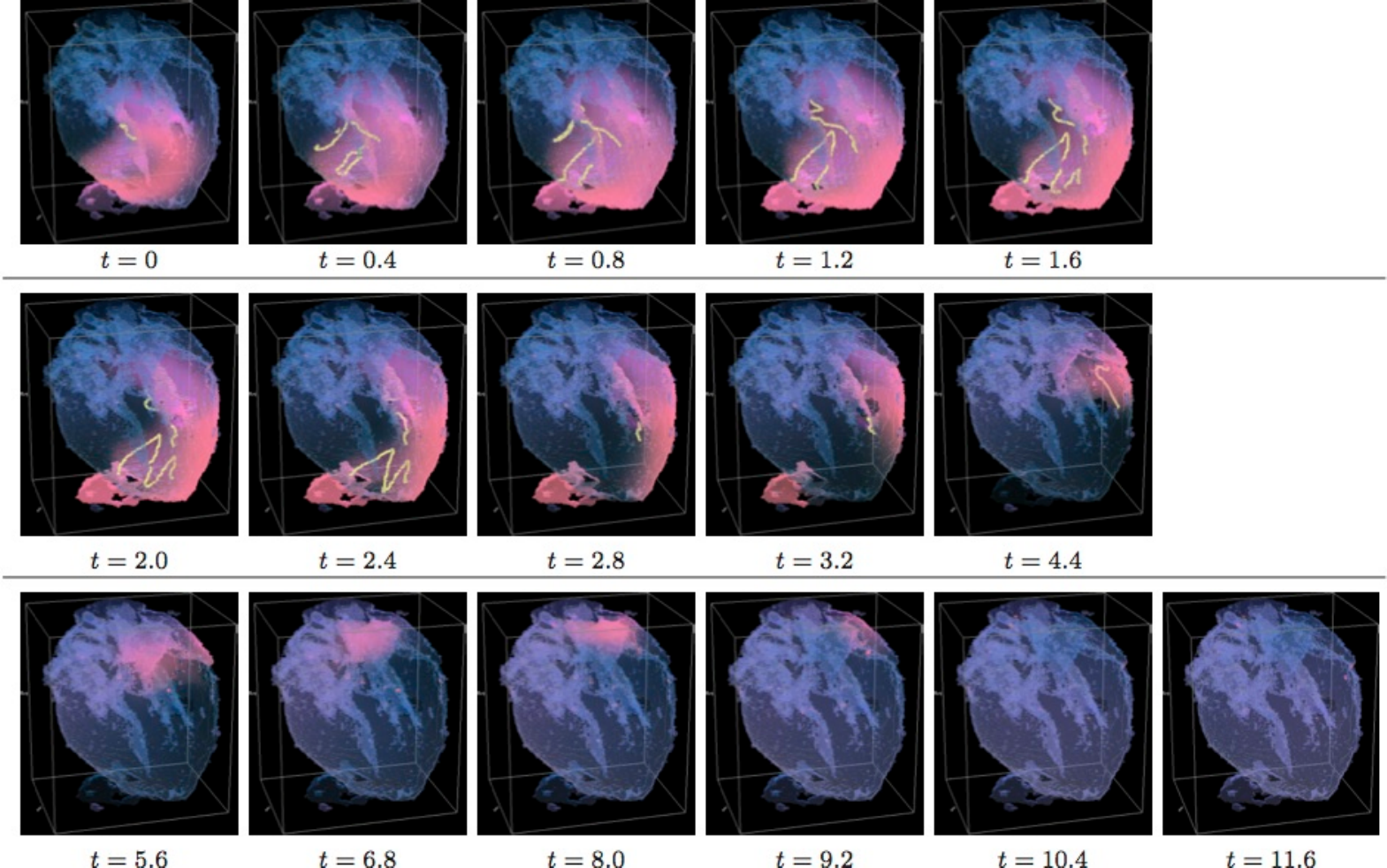}%
\caption{\label{3D_aniso} {\bf Anisotropic whole heart 
    simulation.} The translucent foetal heart is shown
  in blue, excitation front shown in red (see the color box in FIG.~\ref{3D_iso}), the yellow lines are the
  instant organising filaments of the excitation
  vortices; time shown \chg[units]{under each
  panel in time units of Eqs.~(\ref{bc})-(\ref{FHN})}. The anisotropy of the heart causes the fast transient distortion of
the organising filament of the initial excitation vortex and drift towards the inexcitable boundary of the heart, ultimately
resulting in the very fast self termination of the excitation vortex. (Multimedia view) Fig4.mpg.}
\end{figure*}

In the 3D whole heart MRI-based simulations shown in the Fig.~\ref{3D_iso}%
and Fig.~\ref{3D_aniso}, %
a counter-clockwise excitation vortex was initiated by the phase distribution
method~\cite{chaos},  with the initial position of the transmural 
vortex filament (yellow line) at the
prescribed location \chg[identical]{\emph{along the $x$ axis}} at $y_0=40, z_0=60$. 
\chg[identical]{It can be seen in Fig.~\ref{3D_iso} isotropic, and
  Fig.~\ref{3D_aniso} anisotropic %
3D simulations that, at $t=0$, there
was identical initial location of  the filament of the
excitation vortex: that is transmurally, roughly in the middle through the
ventricles of
the heart. } 

Fig.~\ref{3D_iso} shows the \emph{isotropic dynamics} of the excitation vortex,
that is 
with the fiber orientation data ``turned OFF'', so that only the
geometry of the otherwise isotropic homogeneous foetal heart affects the
dynamics of the vortex. It
is known that the chosen FHN kinetics parameter values $\paralp=0.3$, $\parbet=\chg[]{0.71}$,
 $\pargam=0.5$ produce rigidly rotating vortex with the positive filament
 tension~\cite{ft}, which, depending on the topology, either
 collapses or straightens up between the opposite
 boundaries of the excitable medium. \chg[leftover]{\chg[leftover-2]{In the 3D anatomically realistic
 isotropic simulations of the foetal heart, the anatomically
 realistic boundaries of the heart cause drift of the excitation
 vortex, and, depending on the initial position of the vortex filament,
 vortices with the positive filament tension tend to collapse. However,
 there exist initial locations of the excitation vortex, which although
 result in the drift of the vortex, still do not lead to the expected
 collapse of the vortex with positive filament tension. One of
 such outcomes is shown in the Fig.~\ref{3D_iso}. Here, following the geometry of the heart, 
 after a very short transient, the initial vortex filament breaks into the
 two short pieces, each of which finds its own synchronous perpetual
 pathway  in the
 ``isotropic'' foetal
 heart, resulting in the seemingly perpetual cardiac re-entry, which failed to self-terminate within
 the extended simulation time, see Fig.~\ref{3D_iso}.}}

Fig.~\ref{3D_aniso} shows the \emph{anisotropic dynamics} of the excitation vortex,
that is 
with the fiber orientation data ``turned ON'', so that both the
anatomically realistic 
geometry and the anisotropy of the otherwise homogeneous foetal heart
affect the 
dynamics of the initial vortex.

\begin{figure*}[tb]
\includegraphics{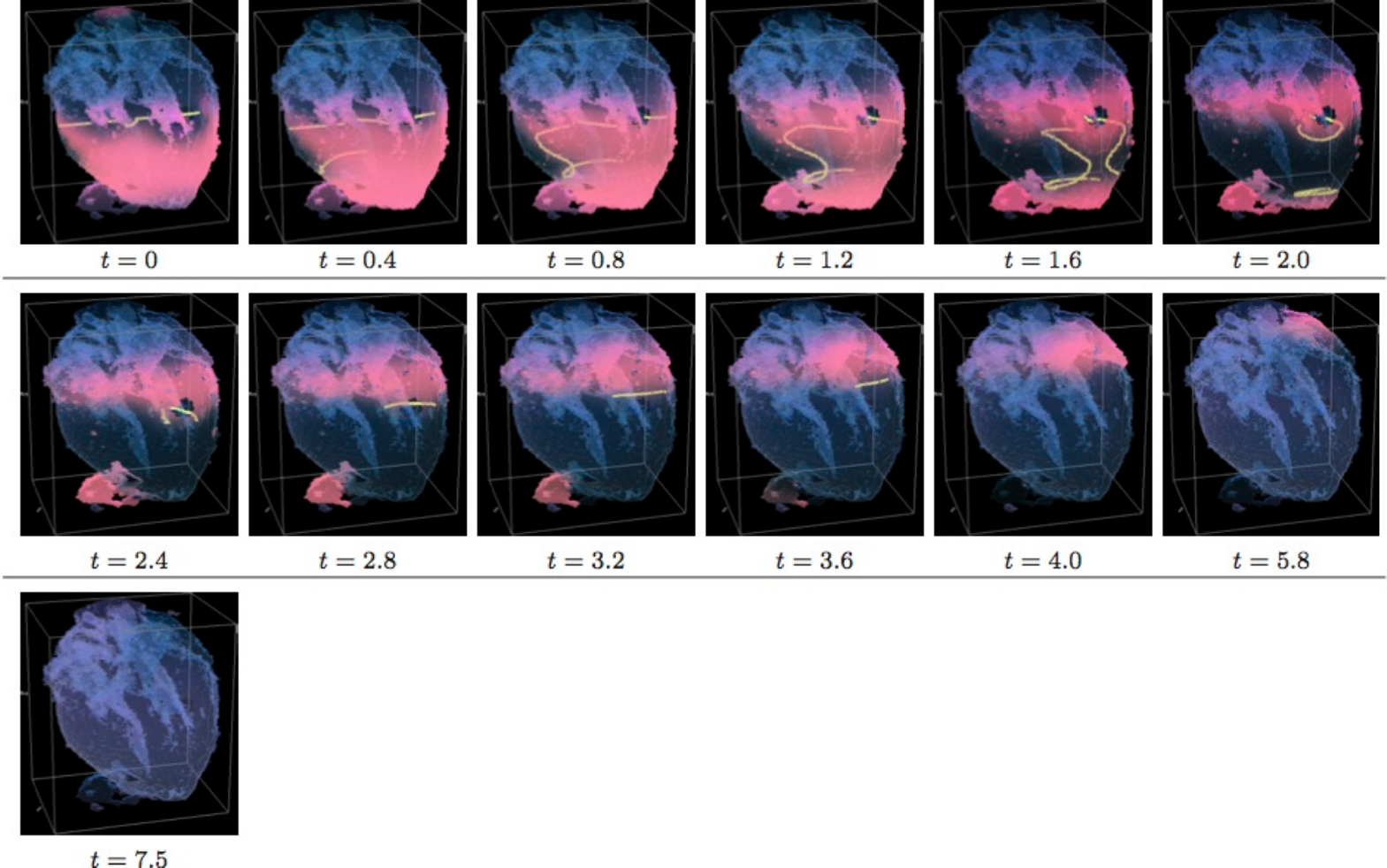}%
\caption{\label{3D_X_iso} {\bf Isotropic whole heart 
    simulation.} The translucent foetal heart is shown
  in blue, excitation front shown in red (see the color box in FIG.~\ref{3D_iso}), the yellow lines are the
  instant organising filaments of the excitation
  vortices; time shown \chg[units]{under each
  panel in time units of Eqs.~(\ref{bc})-(\ref{FHN})}. After a short transient the organising filament of the initial vortex breaks into the
 two pieces each of which fast terminates: one at the base and another
 at the apex of the heart. (Multimedia view) Fig5.mpg.}
\end{figure*}

\begin{figure*}[tb]
\includegraphics{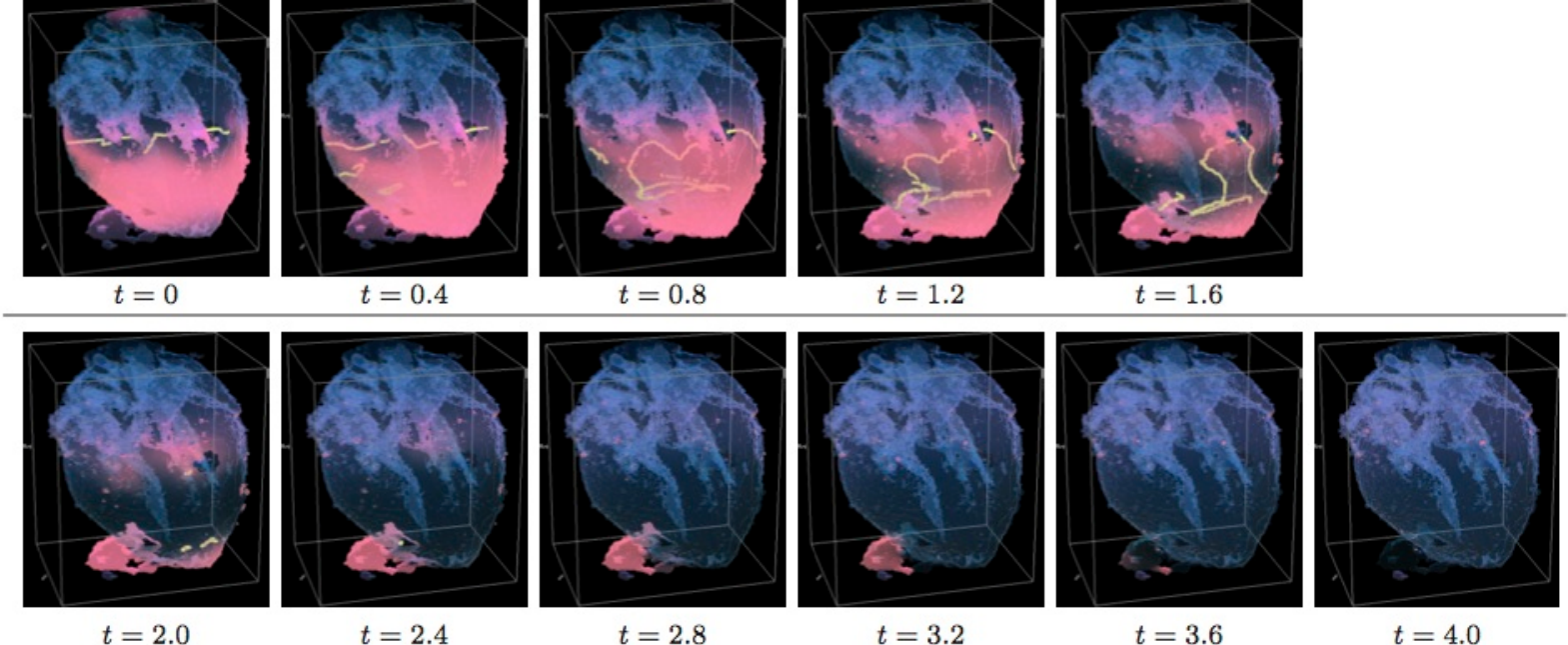}%
\caption{\label{3D_X_aniso} {\bf Anisotropic whole heart 
    simulation.} The translucent foetal heart is shown
  in blue, excitation front shown in red (see the color box in FIG.~\ref{3D_iso}), the yellow lines are the
  instant organising filaments of the excitation
  vortices; time shown \chg[units]{under each
  panel in time units of Eqs.~(\ref{bc})-(\ref{FHN})}. The anisotropy of the heart causes the fast transient distortion of
the organising filament of the initial excitation vortex, followed by
the fast drift and self-termination at the apex of the heart. (Multimedia view) Fig6.mpg.}
\end{figure*}

In the 3D whole heart MRI-based simulations shown in the Fig.~\ref{3D_X_iso}%
and Fig.~\ref{3D_X_aniso}, %
a counter-clockwise excitation vortex was initiated by the phase distribution
method~\cite{chaos},  with the initial position of the transmural 
vortex filament (yellow line) at the
prescribed location \chg[identical]{\emph{along the $y$ axis}} at
$x_0=40, z_0=60$\chg[identical]{, that is \emph{perpendicular} to the
  initial orientation of the vortex filament shown in Fig.~\ref{3D_iso} and Fig.~\ref{3D_aniso}.
It can be seen in Fig.~\ref{3D_X_iso} isotropic, and in Fig.~\ref{3D_X_aniso} %
anisotropic 3D simulations, that at $t=0$, there
was identical intial location of the filament of the
excitation vortex: that is transmurally, roughly in the middle through the
ventricles of
the foetal heart, and \emph{perpendicular} to the
  initial orientation of the vortex filament shown in Fig.~\ref{3D_iso} and Fig.~\ref{3D_aniso}.  }

Fig.~\ref{3D_X_iso} shows the \emph{isotropic dynamics} of the excitation vortex,
that is 
with the fiber orientation data ``turned OFF'', so that only the
geometry of the otherwise isotropic homogeneous foetal heart affects the
dynamics of the vortex. Here, contrary to the expectation for
the positive filament tension vortex to always contract, the
organising filament first
transiently extends intramurally along the tissue walls, before finally breaking up
to the two ring-like pieces, each of which quickly contracts and
terminates at the
opposite base and apex regions of the heart.

Fig.~\ref{3D_X_aniso} shows the \emph{anisotropic dynamics} of the excitation vortex,
that is 
with the fiber orientation data ``turned ON'', so that both the
anatomically realistic 
geometry and the anisotropy of the otherwise homogeneous foetal heart
affect the 
dynamics of the initial vortex leading to its really fast termination
at the apex of the heart.

\chg[leftover]{\chg[leftover-2]{In the raw DT-MRI model simulations
    shown in Fig.~\ref{3D_iso}, Fig.~\ref{3D_aniso},
Fig.~\ref{3D_X_iso}, and Fig.~\ref{3D_X_aniso}, it can be seen that
although the organising filament
of the vortex could not get through into the accidental ``leftover'' piece of
tissue adjacent to the apical region, the piece got activated and might have served as an artificial
``capacitor'' affecting dynamics of the re-entry. In order to
check whether this might be the case, we edited the original raw
DT-MRI model by removing in
the MRI the foreign piece, and repeated the whole heart
isotropic and anisotropic simulations from the same two orthogonal
initial locations of the re-entry, similar to the shown in Fig.~\ref{3D_iso}, Fig.~\ref{3D_aniso},
Fig.~\ref{3D_X_iso}, and Fig.~\ref{3D_X_aniso}.}}

\begin{figure*}[tb]
\includegraphics{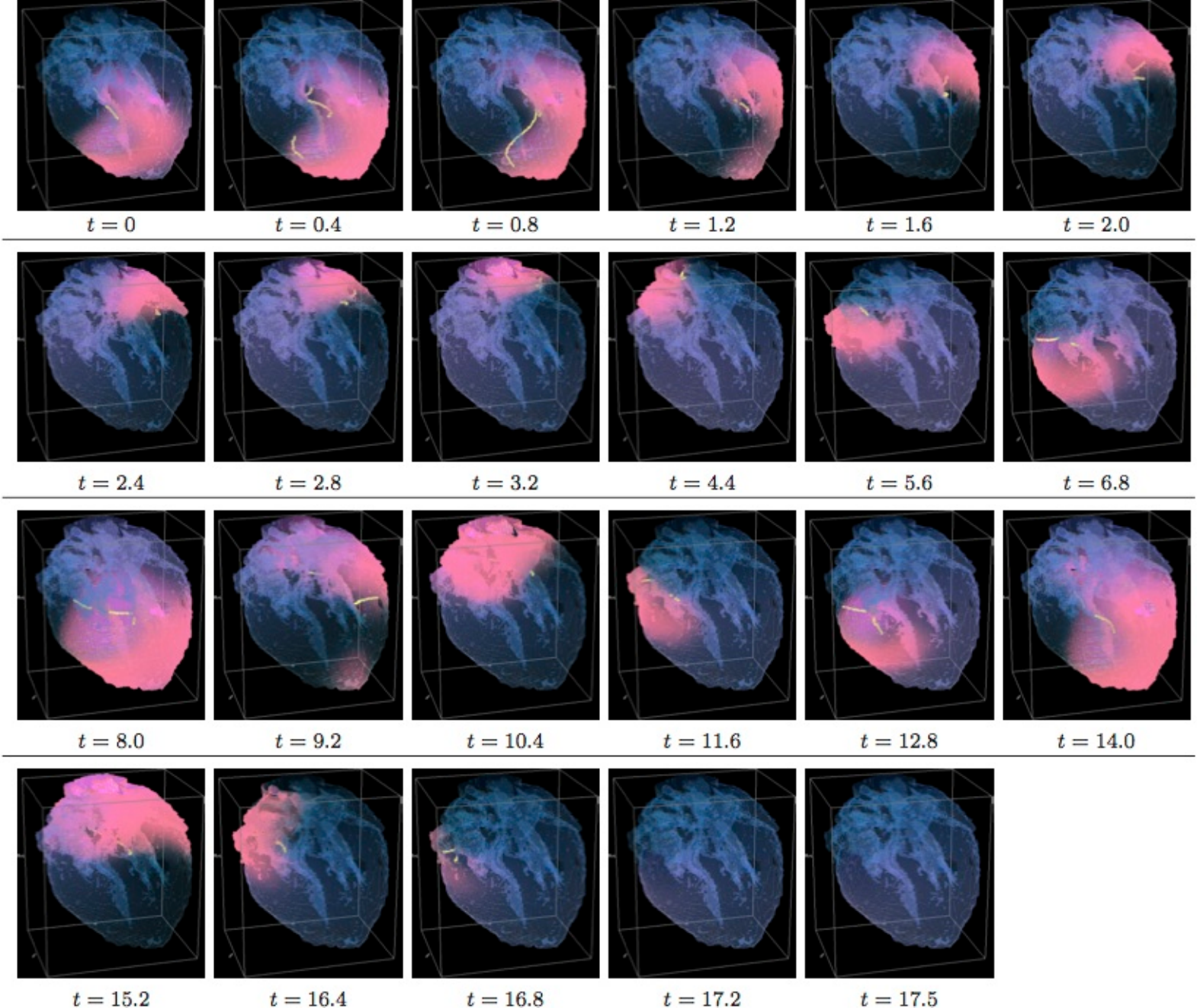}%
\caption{\label{3D_shaven_iso} {\bf Isotropic ``edited'' whole heart 
    simulation.} The translucent foetal heart is shown
  in blue, excitation front shown in red (see the color box in FIG.~\ref{3D_iso}), the yellow lines are the
  instant organising filaments of the excitation
  vortices; time shown \chg[units]{under each
  panel in time units of Eqs.~(\ref{bc})-(\ref{FHN})}. After a short transient the organising filament of the initial vortex breaks into the
 two short pieces each of which finds its own synchronous
 pathway, resulting after a few rotations in the synchronous
 termination of the filaments in the base of the foetal
 heart. (Multimedia view) Fig7.mpg.}
\end{figure*}

\begin{figure*}[tb]
\includegraphics{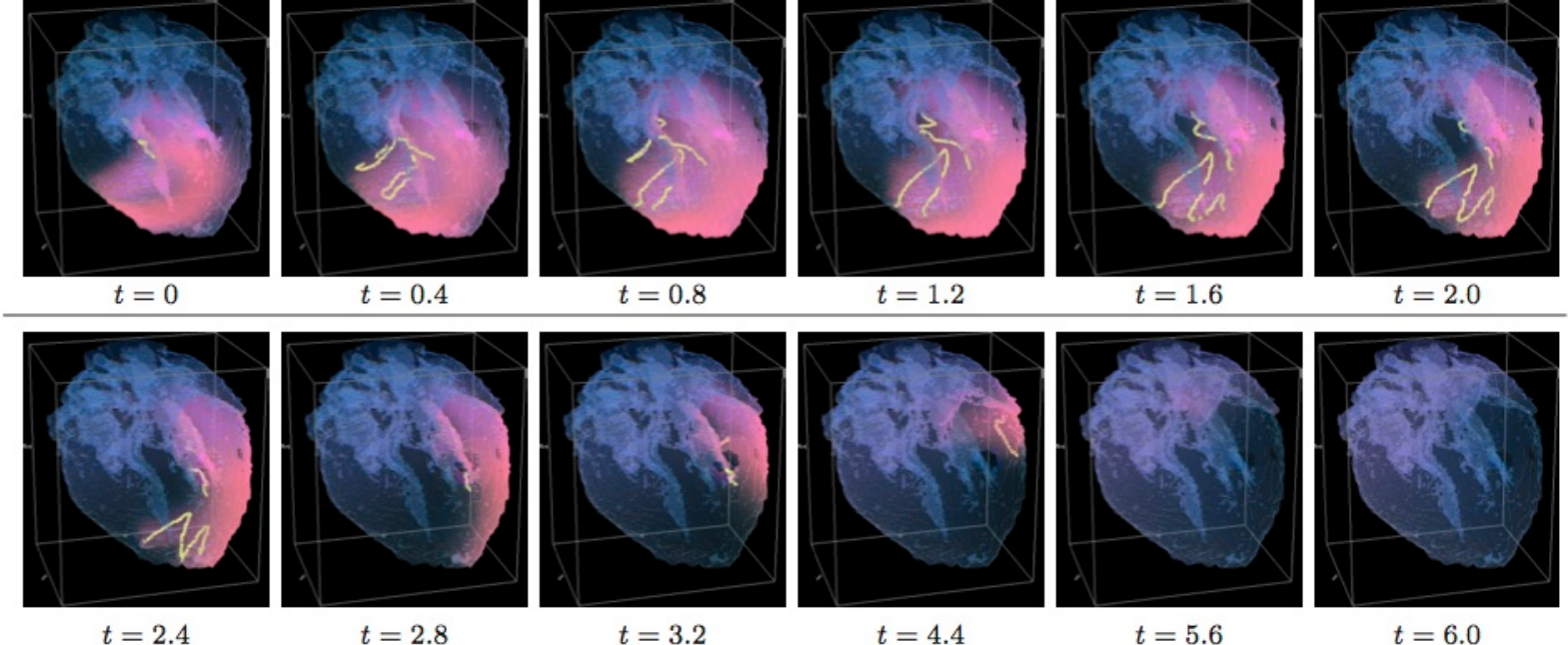}%
\caption{\label{3D_shaven_aniso} {\bf Anisotropic ``edited'' whole heart 
    simulation.} The translucent foetal heart is shown
  in blue, excitation front shown in red (see the color box in FIG.~\ref{3D_iso}), the yellow lines are the
  instant organising filaments of the excitation
  vortices; time shown \chg[units]{under each
  panel in time units of Eqs.~(\ref{bc})-(\ref{FHN})}. The anisotropy of the heart causes the significant transient distortion of
the organising filament of the initial vortex, followed by its fast
drift towards the apex and the ultimate termination before completing
a single rotation. (Multimedia view) Fig8.mpg.}
\end{figure*}

In the 3D whole heart ``edited'' MRI model simulations shown in the Fig.~\ref{3D_shaven_iso}%
and Fig.~\ref{3D_shaven_aniso}, %
a counter-clockwise excitation vortex was initiated by the phase distribution
method~\cite{chaos},  with the initial position of the transmural 
vortex filament (yellow line) at the
prescribed location \chg[identical]{\emph{along the $x$ axis}} at $y_0=40, z_0=60$. 
\chg[identical]{It can be seen in Fig.~\ref{3D_shaven_iso} isotropic, and in Fig.~\ref{3D_shaven_aniso} %
anisotropic 3D simulations, that, at $t=0$, there
was identical initial location of  the filament of the
excitation vortex: that is transmurally, roughly in the middle through the
ventricles of
the foetal heart, similar to the initial location of the vortex
filament in the raw DT-MRI simulations shown in Fig.~\ref{3D_iso} and Fig.~\ref{3D_aniso} .  }

Fig.~\ref{3D_shaven_iso} shows the \emph{isotropic dynamics} of the excitation vortex,
that is 
with the fiber orientation data ``turned OFF'', so that only the
geometry of the otherwise isotropic homogeneous foetal heart affects the
dynamics of the vortex. Here, following the geometry of the heart, 
the organising filament of the initial vortex \chg[identical]{ also breaks into the two
short pieces, each of which also finds its own
  synchronous pathway similar to the beginning of the raw DT-MRI
  simulation shown in Fig.~\ref{3D_iso}. However, this time, after a few
  rotations, the two re-entries find their end in their also almost synchronous termination of the
filaments in the base region of the foetal heart, see Fig.~\ref{3D_shaven_iso}.}

Fig.~\ref{3D_shaven_aniso} shows the \emph{anisotropic dynamics} of the excitation vortex,
that is 
with the fiber orientation data ``turned ON'', so that both the
anatomically realistic 
geometry and the anisotropy of the otherwise homogeneous foetal heart
affect the 
dynamics of the initial vortex. Here, the anisotropy of the heart
\chg[identical]{ also causes the significant transient distortion of
the organising filament of the initial vortex, followed by its fast
drift towards the apex, and the ultimate termination at the AV border
before a completion of 
a single rotation, very similar to the raw DT-MRI
  simulation shown in Fig.~\ref{3D_aniso}. However, this time, without the
  ``leftover'' piece ``incidental capcitor'' effect, there is
  just a bit faster repolarisation of the whole heart than it was in
  the presence of the ``incidental capcitor'' in the raw DT-MRI
  simulation shown in Fig.~\ref{3D_aniso}. \chg[quantifying]{For the
    comparison of the re-entry termination times, and the whole heart recovery
    times, see FIG.~\ref{Table1}}. } 

\begin{figure*}[tb]
\includegraphics{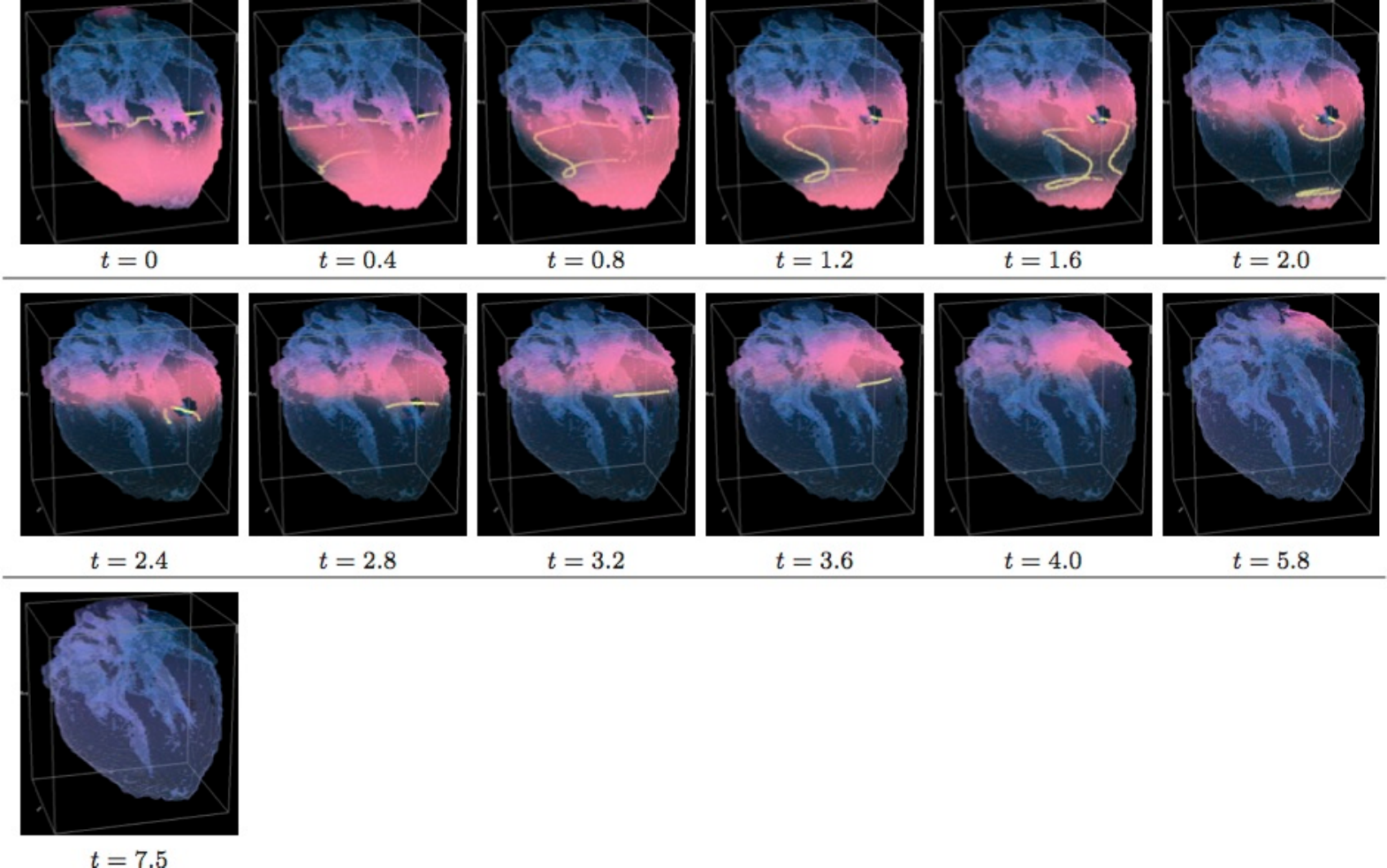}%
\caption{\label{3D_X_shaven_iso} {\bf Isotropic ``edited'' whole heart 
    simulation.} The translucent foetal heart is shown
  in blue, excitation front shown in red (see the color box in FIG.~\ref{3D_iso}), the yellow lines are the
  instant organising filaments of the excitation
  vortices; time shown \chg[units]{under each
  panel in time units of Eqs.~(\ref{bc})-(\ref{FHN})}. After a short transient the organising filament of the initial vortex breaks into the
 two pieces each of which fast terminates: one at the base and another
 at the apex of the heart. (Multimedia view) Fig9.mpg.}
\end{figure*}

\begin{figure*}[tb]
\includegraphics{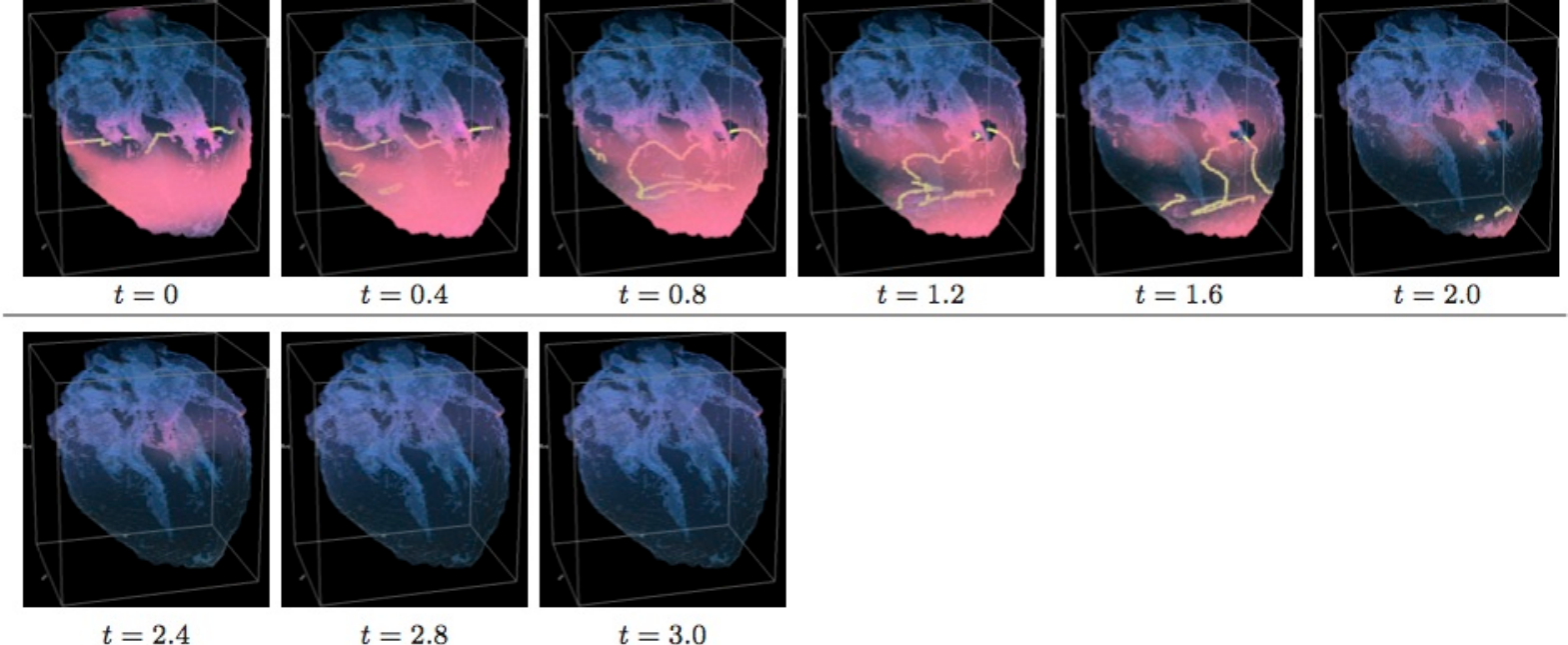}%
\caption{\label{3D_X_shaven_aniso} {\bf Anisotropic ``edited'' whole heart 
    simulation.} The translucent foetal heart is shown
  in blue, excitation front shown in red (see the color box in FIG.~\ref{3D_iso}), the yellow lines are the
  instant organising filaments of the excitation
  vortices; time shown \chg[units]{under each
  panel in time units of Eqs.~(\ref{bc})-(\ref{FHN})}. The anisotropy of the heart causes the fast significant transient distortion of
the organising filament of the initial excitation vortex, followed by
the fast drift towards the apex and ultimate termination before the
first rotaion has ever started. (Multimedia view) Fig10.mpg.}
\end{figure*}

In the 3D whole heart ``edited'' MRI model simulations shown in the Fig.~\ref{3D_X_shaven_iso}%
and Fig.~\ref{3D_X_shaven_aniso}, %
a counter-clockwise excitation vortex was initiated by the phase distribution
method~\cite{chaos},  with the initial position of the transmural 
vortex filament (yellow line) at the
prescribed location \chg[identical]{\emph{along the $y$ axis}} at $x_0=40, z_0=60$. 
\chg[identical]{It can be seen in Fig.~\ref{3D_X_shaven_iso} isotropic, and in Fig.~\ref{3D_X_shaven_aniso} %
anisotropic 3D simulations, that at $t=0$, there
was the identical initial location of  the filament of the
excitation vortex: that is transmurally, roughly in the middle through the
ventricles of
the foetal heart, perpendicular to the initial location of the vortex
filament in the ``edited'' MRI simulations shown in Fig.~\ref{3D_shaven_iso} and
Fig.~\ref{3D_shaven_aniso}, and similar to the initial location of the vortex
filament in the raw DT-MRI simulations shown in Fig.~\ref{3D_X_iso} and Fig.~\ref{3D_X_aniso} .  }

\chg[identical]{Fig.~\ref{3D_X_shaven_iso} shows the \emph{isotropic dynamics} of the excitation vortex,
that is 
with the fiber orientation data ``turned OFF'', so that only the
geometry of the otherwise isotropic homogeneous foetal heart affects the
dynamics of the vortex. Here, again contrary to the expectation for
a positive filament tension vortex to always contract, the
organising filament first
transiently extends intramurally before \chg[]{breaking up}
into the two ring-like pieces, each of which quickly contracts and
terminates at the
opposite base and apex regions of the heart,
identical to what can be seen in the raw DT-MRI
    simulation shown in Fig.~\ref{3D_X_iso}. So that, this time,
    for this particular orientation of the initial re-entry, the ``leftover''
    tissue ``incidental capacitor'' effect does not seem to play any
    role in the outcomes of the isotropic ``heart geometry only'' raw
    DT-MRI simulations shown in Fig.~\ref{3D_X_iso} as opoosed to
    the outcome of the identical initial re-entry location in the 
    ``edited'' MRI simulations shown in Fig.~\ref{3D_X_shaven_iso}. }

Fig.~\ref{3D_X_shaven_aniso} shows the \emph{anisotropic dynamics} of the excitation vortex,
that is 
with the fiber orientation data ``turned ON'', so that both the
anatomically realistic 
geometry and the anisotropy of the otherwise homogeneous foetal heart
affect the 
dynamics of the initial vortex\chg[identical]{, which, in the absence of the
``incidental capacitor'' effect, results in the fastest possible
termination of the re-entry 
at the apex of the heart, before the vortex first rotation ever
started. The re-entry termination time here is more than twice shorter than in the
raw and ``edited'' MRI isotropic simulations shown in
Fig.~\ref{3D_X_iso} and Fig.~\ref{3D_X_shaven_iso}, shorter than in
the analogous simulation with the ``incidental capacitor'' effect shown in the Fig.~\ref{3D_X_aniso}, and times shorter
than in any of the simulations of the re-entry with the perpendicular initial location shown
in the Fig.~\ref{3D_iso}, Fig.~\ref{3D_aniso},
Fig.~\ref{3D_shaven_iso}, and Fig.~\ref{3D_shaven_aniso} .}  

In Fig~\ref{Table1}, we have summarized the results of the simulations
shown in Fig.~\ref{3D_iso}, Fig.~\ref{3D_aniso},
Fig.~\ref{3D_X_iso}, Fig.~\ref{3D_X_aniso}, Fig.~\ref{3D_shaven_iso}, Fig.~\ref{3D_shaven_aniso},
Fig.~\ref{3D_X_shaven_iso}, and Fig.~\ref{3D_X_shaven_aniso}, with the
re-entry termination time shown in arbitrary time units
  under each respected whole heart model and initiation cite
  panel.  It can be seen that the realistic anisotropy of the heart
  causes at least twice faster termination of re-entry.
It also can be seen that indeed the present in the raw DT-MRI model
leftover piece of tissue connected to
the apical region of the heart has served as an artificial
``capacitor'' affecting the dynamics of the re-entry, and
significantly prolongated life time of the re-entry initiated at
particular locations/orientation respective to the ``capacitor''. 

Finally, the 3D anatomically realistic simulations of the
 foetal heart show that 
the realistic anisotropy of the heart causes the fast transient distortion of
the vortex filament, and the typical fast drift towards the apex area
of the inexcitable boundary of the heart, which ultimately
results in the fast self-termination of the excitation vortex, see
Figs.~\ref{3D_iso}-\ref{3D_X_shaven_aniso} and the corresponding movies
 in the Supplementary Material section~\ref{Suppl}.

\section{\label{Discussion} DISCUSSION AND FUTURE DIRECTIONS}
\begin{figure*}[tb]
\includegraphics{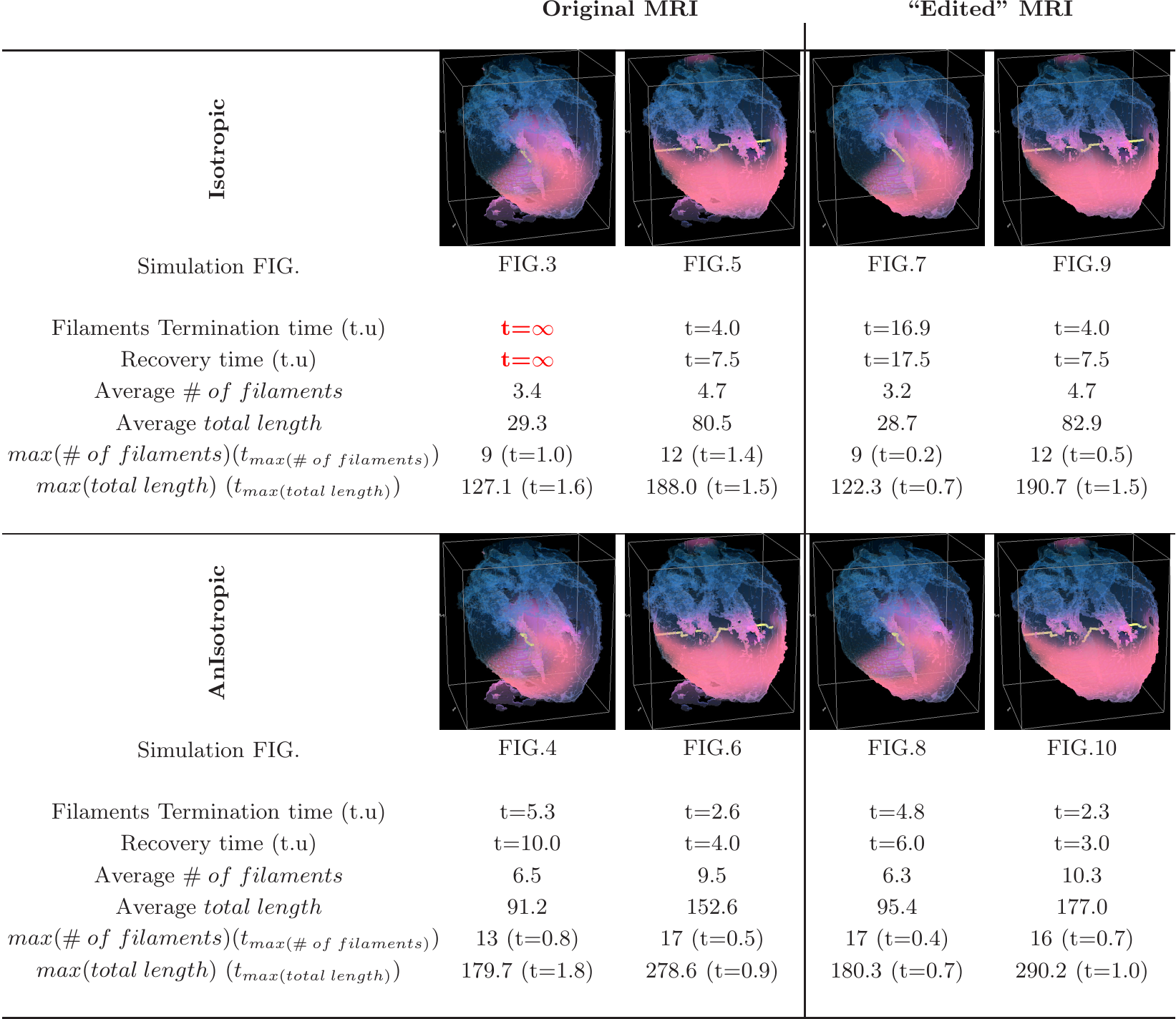}%
\caption{\label{Table1} {\bf Whole heart simulation: re-entry termination times.} The translucent foetal heart is shown
  in blue, excitation front shown in red (see the color box in FIG.~\ref{3D_iso}), the yellow lines are the
  instant organising filaments of the excitation
  vortices. \chg[]{Re-entry self-termination time \chg[units]{in time units of Eqs.~(\ref{bc})-(\ref{FHN})} is shown under each
  simulation FIG.\ref{3D_iso}--FIG.\ref{3D_X_shaven_aniso}
   initiation panel.}
 \chg[leftover]{\chg[leftover-2]{Comparison of the respective isotropic (top row) vs
   anisotropic (bottom row) simulations shows that, regardless of with
   or without the ``leftover'' piece, anisotropy results in  
   faster termination of re-entry, and at least twice shorter recovery
   time. 
Respective comparison of the original MRI with the corresponding ``edited out
leftover'' simulations shows that the leftover ``incidental capacitor'' effect,
depending of the re-entry location/orientation with respect to the
``incidental capacitor'' own location/orientation, might significantly
prolongate cardiac re-entry life time.} \chg[quantifying]{The
  bigger number and the total length of the filaments tend to correlate
  with the faster termination of re-entry, though these fail to
  identify persistent re-entry in FIG.~\ref{3D_iso}
  simulation. }}
}
\end{figure*}

\begin{figure*}[tb]
\includegraphics{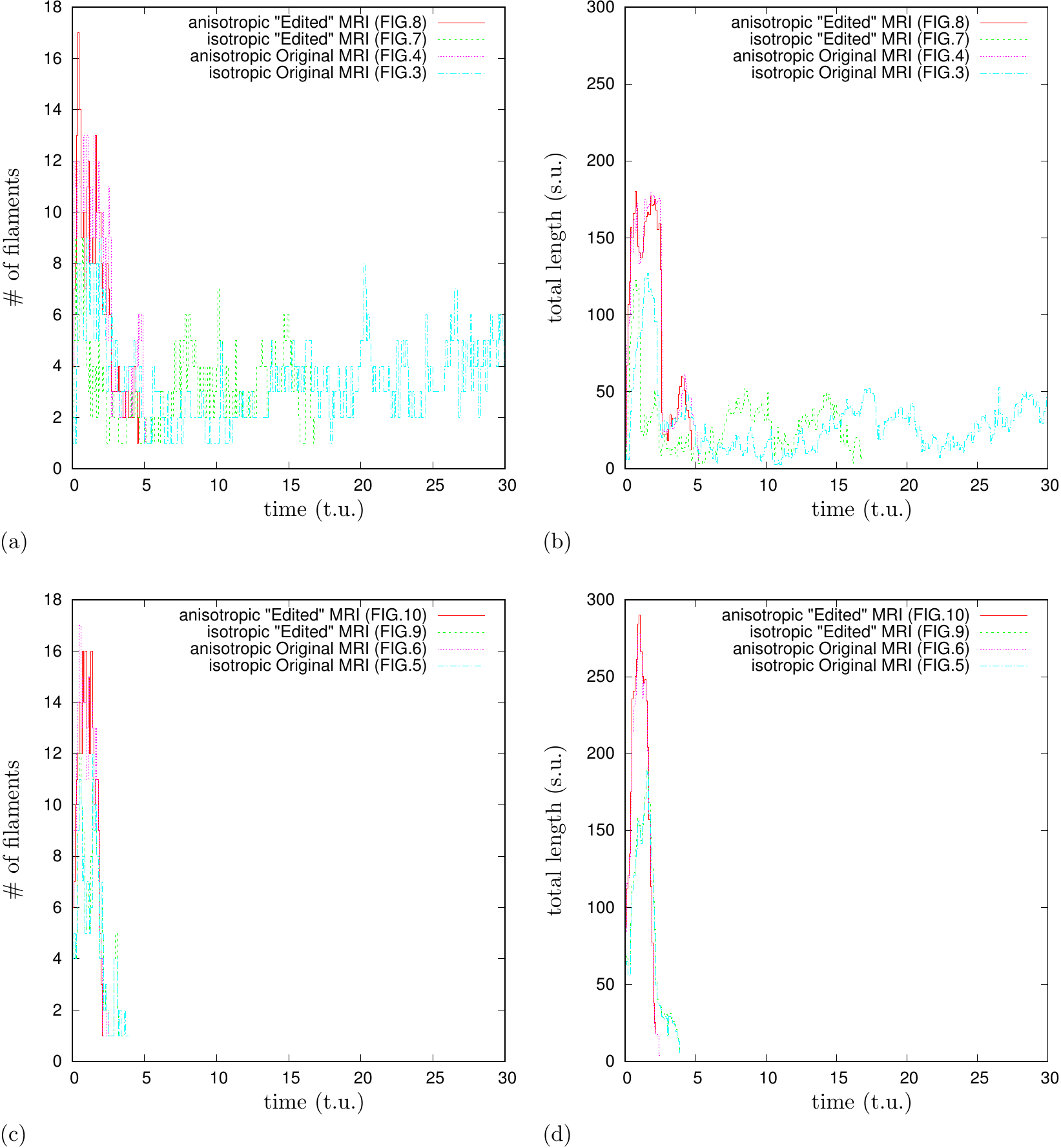}%
\caption{\label{quant} \chg[quantifying]{{\bf Whole heart simulation: time course of the
    number of filaments \#, and of the total length of the filaments},
  time and the total length of the filaments shown \chg[units]{in the time
    and space units of Eqs.~(\ref{bc})-(\ref{FHN})}.  {\bf Initial position of the 
vortex filament \emph{along the $x$ axis}}, FIG.~\ref{3D_iso}, 
FIG.~\ref{3D_aniso}, FIG.~\ref{3D_shaven_iso}, and 
FIG.~\ref{3D_shaven_aniso} simulations: {\bf (a)} time course of the
    number of filaments \# ; {\bf (b)} time course of the
    total length of the filaments. 
 {\bf Initial position of the 
vortex filament \emph{along the $y$ axis}}, FIG.~\ref{3D_X_iso},
FIG.~\ref{3D_X_aniso}, FIG.~\ref{3D_X_shaven_iso}, and 
FIG.~\ref{3D_X_shaven_aniso} simulations:
{\bf (c)} time course of the
    number of filaments \# ; {\bf (d)}  time course of the
    total length of the filaments. 
Anisotropy
  increases the transient number and the transient total length of the filaments. The
  bigger transient number and the total length of the filaments tend to correlate
  with the faster termination of re-entry. The biggest transient total
  lengths of the filaments was in case of the re-entry initiated
  \emph{along the $y$ axis}, panel {\bf (d)}, which ensured its 
fastest termination. It can be seen from FIG.~\ref{3D_X_iso},
FIG.~\ref{3D_X_aniso}, FIG.~\ref{3D_X_shaven_iso}, and 
FIG.~\ref{3D_X_shaven_aniso}, that the initial position of the
filament \emph{along the $y$ axis} allowed it to grow intramurally,
thus maximally increasing the transient total
length of the filaments, and speeding up their termination.  
}}
\end{figure*}

 Although the role of heart anatomy and anisotropy in the origin
and sustainability of cardiac
arrhythmias has been appreciated for a long time, the experimental
evidence capable to clarify the detail of the effects of the heart anatomy on
the persistent 
cardiac arrhythmias and fibrillation 
are limited. 
\chg[mechanistic]{\chg[incidental]{
In particular, the theoretically plausible hypothesis that the anisotropic discontinuities in
the heart might be a source of rise for cardiac re-entry due to the abrupt change in
conduction velocity and wavefront curvature\cite{Fenton-Karma-1998, 
  Spach-CircRes-2001, Smaill-etal-2004} was in controversy with the
observation that the transmural fiber
arrangement, including the range of transmural change in fiber angle in ventricular wall, although varied between  
species\cite[p.~173]{Hunter-etal-CompBiolOfHeart}, was consistent within a
species. So that the question was that, if the pro-arrhythmic mechanism of cardiac re-entry
initiation by the anisotropic discontinuities in a
heart\cite{Fenton-Karma-1998, Spach-CircRes-2001, Smaill-etal-2004}
was correct, what would then have been a reason for the
consistent structure~\cite[p.~173]{Hunter-etal-CompBiolOfHeart} of the anisotropic discontinuities in healthy mammalian hearts.}
The} combination of the High Performance Computing with the high-resolution DT-MRI
based anatomy models of the heart 
allows anatomically realistic \emph{in-silico}
testing of the effects of individual heart anatomy and anisotropy on the cardiac
re-entry dynamics~\cite{Kharche-etal-2015-BMRI, Kharche-etal-2015-LNCS,bbx-2017-PONE}.
In this paper, for the first time, we present the anatomy and myofiber structure  
realistic computer simulation study of the cardiac re-entry dynamics in
the DT-MRI based  model of the human foetal
heart~\cite{Pervolaraki-etal-2013}. 

The comparative isotropic \emph{vs}
anisotropic simulation of the otherwise homogeneous foetal heart shows that,
in the 2D slice of the heart, the realistic fiber anisotropy might
change the re-entry dynamics
from pinning to the sharp end of the septum cuneiform opening, Fig.~\ref{slice_fig}(a), into
a fast anatomical re-entry around the opening,
Fig.~\ref{slice_fig}(b). \chg[segmentation-2]{Because of the 2D re-entry pinning to either
the sharp end of the septum opening in the isotropic
simulation in Fig.~\ref{slice_fig}(a), or to the whole septum
opening as an anatomical re-entry in the anisotropic simulation in
Fig.~\ref{slice_fig}(b), despite of the only basic segmentation of the
MRI model into the tissue/not tissue points, and the
ventricles not being isolated from the atria, the tip of the re-entry never got from the
ventricles into the atria, Fig.~\ref{slice_fig}. }
\chg[]{Although, from the cardiac
physiology point of view, the 
\chg[segmentation]{\chg[segmentation-2]{ \chg[epi/endo]{
only basic segmentation of the raw DT-MRI data~\cite{Pervolaraki-etal-2013} 
into the tissue/non-tissue pixels}
might be seen as}} 
a major limitation of the study, from the non-linear science point
of view, the use of the raw MRI data as an
example of a nature provided medium to study a re-entry dynamics gives
an important insight into the pure anatomy induced drift in an
otherwise homogeneous 2D medium, and into the 
possibility of pinning
of the re-entry not to a major blood vessel but to a sharp end of an
anatomical openning~\cite{Biktasheva-etal-2015-PRL}; and into that a real fiber anisotropy is capable to
turn the pinned re-entry into an anatomical one.} Importantly though, the 2D simulations in Fig.~\ref{slice_fig} are an
important step to highlight the role and the necessity of the whole heart
structure in the re-entry dynamics and self-termination.

\chg[incidental]{In the 3D
DT-MRI based \emph{isotropic} model of the foetal heart, depending on
the initial location/orientation of the organising filament of the
excitation vortex, the geometry of the foetal 
heart might
sustain perpetual cardiac re-entry even with a positive filament
tension, Fig.~\ref{3D_iso}. However, if the same positive filament tension vortex is
initiated at the exactly same
location/orientation and in the same anatomical environment in the full \emph{anisotropic} 3D DT-MRI based model of
the heart, the realistic fiber structure of the foetal heart facilitates fast self-termination
of cardiac re-entry, Fig.~\ref{3D_aniso}. }

\chg[segmentation-2]{ From the respective
comparison of the ``isotropic vs anisotropic'' simulations in FIG.~\ref{3D_iso} vs
FIG.~\ref{3D_aniso}, and FIG.~\ref{3D_shaven_iso} vs
FIG.~\ref{3D_shaven_aniso}, it can be seen that, whereas the re-entry
filaments were capable to
penetrate from the ventricles to atria in the isotropic simulations
shown in
FIG.~\ref{3D_iso}  and FIG.~\ref{3D_shaven_iso}, the abrupt change in
the fiber angles between the atria and the ventricles, which can be
seen in FIG.~\ref{HF_slice_x_63}, did not allow the re-entry
filaments to
get from the ventricles to atria in the anisotropc simulations shown in
FIG.~\ref{3D_aniso}  and FIG.~\ref{3D_shaven_aniso}, so that the ventricles'
anisotropy could complete the speedy elimination of the re-entry within its
single rotation. }

\chg[leftover]{\chg[leftover-2]{\chg[quantifying]{The comparison of
      the re-entry termination times in the raw DT-MRI data model whole heart
  simulations shown in Fig.~\ref{3D_iso}, Fig.~\ref{3D_aniso},
Fig.~\ref{3D_X_iso}, and Fig.~\ref{3D_X_aniso}, with the corresponding
series of the ``edited'' MRI model whole heart simulations shown in Fig.~\ref{3D_shaven_iso}, Fig.~\ref{3D_shaven_aniso},
Fig.~\ref{3D_X_shaven_iso}, and Fig.~\ref{3D_X_shaven_aniso}, showed
that, although the 
filament of the re-entry never got through into the small piece
of excitable 
tissue accidentally adjacent to the apical region of the heart, the
adjacent tissue  served as a ``capacitor'' 
significantly prolongating the life time of the re-entry initiated at
a particular location/orientation respective to the 
``capacitor's'' own location/orientation. See for the quantitative
comparison of the re-entry termination times Fig.~\ref{Table1} and
Fig.~\ref{quant}, where the
  bigger number and the total length of the filaments tend to correlate
  with the faster termination of re-entry, though these fail to
  identify the persistent re-entry in FIG.~\ref{3D_iso}
  simulation. }

The ``isotropic vs anisotropic'' comparison of re-entry
self-termination time in both the original raw
DT-MRI simulations series, and in the ``edited'' MRI
whole heart simulations, confirmed that the real anisotropy of the heart speeds up
cardiac re-entry self-termination. \chg[quantifying]{The re-entry self-termination
times provided for the summarised comparison in 
Fig.~\ref{Table1} and  Fig.~\ref{quant}, show that, regardless of with or without the ``leftover'' piece
adjacent to the apex, the anisotropy of the heart speeds up
cardiac re-entry self-termination. Fig.~\ref{quant} 
shows that anisotropy
  increases the transient number and the transient total length of the filaments. The
  bigger transient number and the total length of the filaments tend to correlate
  with the faster termination of re-entry. The biggest transient total
  length of the filaments was in case of the re-entry initiated
  \emph{along the $y$ axis}, see panel {\bf (d)} in Fig.~\ref{quant},
  which ensured the re-entry 
fastest termination. It can be seen from FIG.~\ref{3D_X_iso},
FIG.~\ref{3D_X_aniso}, FIG.~\ref{3D_X_shaven_iso}, and 
FIG.~\ref{3D_X_shaven_aniso}, that the initial position of the
filament \emph{along the $y$ axis} allowed the filament to grow intramurally,
thus maximally increasing the transient total
length of the filaments, and speeding up their termination. }

The simulations with the
``edited'' MRI image of thus completely 
isolated heart, Fig.~\ref{3D_shaven_iso}, Fig.~\ref{3D_shaven_aniso},
Fig.~\ref{3D_X_shaven_iso}, and Fig.~\ref{3D_X_shaven_aniso}, in
comparison with the original DT-MRI model simulations, Fig.~\ref{3D_iso}, Fig.~\ref{3D_aniso},
Fig.~\ref{3D_X_iso}, Fig.~\ref{3D_X_aniso}, provide an
important new biological insight into the problem of cardiac re-entry
dynamics. 
Namely, that an excitable tissue accidentally
adjacent to the heart might serve as a capacitor capable to prolongate
time of cardiac re-entry self-termination, see for the respective
comparison the simulations in Fig.~\ref{3D_iso} against Fig.~\ref{3D_shaven_iso},
Fig.~\ref{3D_aniso} against  Fig.~\ref{3D_shaven_aniso},
Fig.~\ref{3D_X_iso} against Fig.~\ref{3D_X_shaven_iso},
and Fig.~\ref{3D_X_aniso} against Fig.~\ref{3D_X_shaven_aniso}, all also 
summarised in Fig.~\ref{Table1}. The latter suggests a possible new
mechanism for persistent cardiac re-entry. So that if, apart from the major blood vessels
normally adjacent to the heart in vivo and affecting
re-entry dynamics, there were also an accidental
``touching'' of the heart by an adjacent excitable tissue, for example, due to the change of posture in the night
sleep, the ``incidental capacitor'' effect could prolongate the
time of cardiac re-entry self-termination, or indeed failure to
self-terminate, which could be an explanation to 
the elusive and difficult to reporduce but statistically salient data
for longer episodes of arrhythmias reported in the night ECGs as opposed to
the on average shorter arrhythmias in the day time ECGs.
Although our simulations using the original raw DT-MRI data with the
small piece of the foreign leftover tissue, could have been seen a limitation of the
study, 
the real heart in vivo does not exist in complete isolation from the main blood vessels
and other neighboring tissues. So, we believe that our ``incidental'' leftover tissue results
only once more confirm the importance and the necessity of taking
into account the real anatomical settings and surrounding of the heart for the
full appreciation of cardiac re-entry dynamics.
}}

The \bbx ~DT-MRI based ~\emph{in-silico} model comparative study
confirms the cardiac anatomy and anisotropy functional effect on cardiac re-entry
sustainability as opposed to its self termination, the  pinning of the re-entry to anatomical features,
its transformation from pinned to
anatomical re-entry, and the re-entry self-termination caused by the 
anisotropy of the tissue.  

One of the limitations of the present study is the use of the simplified
Fitz-Hugh-Nagumo~\cite{Winfree-1991} excitation model Eq.~(\ref{FHN}).
The simplified FHN model with the excitation kinetics parameters $\paralp=0.3$, $\parbet=\chg[]{0.71}$,
 $\pargam=0.5$, which, in an infinite homogeneous isotropic excitable medium, supports a rigidly rotating vortex with positive filament
 tension~\cite{ft}, was chosen for this study in order to eliminate
 the effects of realistic cell excitation kinetics, such as \emph{e.g.}
 meander~\cite{Winfree-1991}, alternans\cite{Karma_Chaos1994}, negative filament tension\cite{ft},
 etc., in order to enhance and highlight the pure effects of the heart
 anatomy and anisotropy on the cardiac re-entry outcome. The realistic cell excitation models should
 be used in the future studies, in order to clarify the particular
 effects and interplay of the cell excitation kinetics with the heart anatomy
 and anisotropy.    

As it can be seen from Fig.~\ref{HF_slice_x_63} (\chg[color-coding]{for the
      color-encoded fractional anisotropy (FA) and for the
      color-encoded all the three components of
      the fiber angles see Figure 4 in Pervolaraki et al~\cite[p.~5]{Pervolaraki-etal-2013}}), formation of the fiber structure at the epicardium and endocardium is not completed yet
in the foetal heart, so
that only the already formed intramural laminar structure of the
fibers can affect the
dynamics of cardiac re-entry. Although the use of the not fully formed foetal heart can be seen as a
limitation of the study, on the other hand, it may be said that the chaotic epicardium and
endocardium fiber orientation prevents the foetal heart re-entry from
pinning to the fine anatomical features which were yet to be
developed at the fully
  formed~\cite{Pervolaraki-etal-2013} endocardium  later on.  
\chg[incidental]{The possible differences in the anatomy and fiber
  structure between the foetal heart used here and fully formed/adult
  hearts in general, could have seriously affected the simulations,
such as in the case of \emph{e.g}.
reported pinning of cardiac re-entry to the junction of pectinate
muscles with crystae terminalis in adult human
atrium~\cite{Wu-etal-1998-CR,Yamazaki-etal-2012-CVR,Kharche-etal-2015-BMRI}. 
That is, although it is possible to initiate a cardiac re-entry in the
tiny $1.4g$ (at $143$ DGA) foetal heart~\cite{Pervolaraki-etal-2013},
the already formed intramural laminar fiber anisotropy of the foetal heart facilitates
the re-entry  self-termination, Fig.~\ref{Table1}. With the hindsight of the present
study, in a
fully formed adult heart, because of the presence of the pinning opportunities provided by the
endocardium anatomical features~\cite{Wu-etal-1998-CR,Yamazaki-etal-2012-CVR,Kharche-etal-2015-BMRI}, there must exist additional mechanisms to
facilitate cardiac re-entry self-termination~\cite{Clayton-etal-1993}.}

The most serious
limitation of the study is that 
\chg[segmentation]{\chg[segmentation-2]{\chg[epi/endo]{ only the basic segmentation of 
the raw DT-MRI data~\cite{Pervolaraki-etal-2013} 
into the tissue/non-tissue pixels 
based on the MRI luminosity threshold,
and only the primary
eigenvalues of fibres orientation, were taken into account in the
\bbx~\cite{bbx-2017-PONE}  
computer simulation of the cardiac re-entry dynamics}.
Further levels of the model
segmentation, in order to take into account e.g. the heart collagen
skeleton, isolation of ventricles from atria, etc.,
will inevitably change the outcome of the re-entry, by adding the
electrically impermeable barriers to cardiac
re-entry. Currently, this further}} segmentation is added into DT-MRI based models via
complex rule based image
post-processing~\cite{Lombaert-etal-2012,Gahm-etal-2013}, which not
only limits
the available segmented DT-MRI cardiac anatomy data sets, but also inevitably brings in an artificial assumption/limitation element into these
models. From the non-linear science point
of view\chg[]{, which} we have persued in this initial study, the rationale was to use the raw DT-MRI data as an
example of a nature provided medium to study a re-entry dynamics. In the future,
the multichannel computer tomography might offer an automatic tissue
segmentation, so that the
\chg[segmentation]{\chg[segmentation-2]{multi-level}} segmented DT-MRI based heart anatomy models
might become more available, and be used in the
\bbx~\cite{bbx-2017-PONE} anatomically and biophysically realistic simulations of cardiac re-entry
dynamics. 

Finally, we believe \chg[]{that} a simple ``mechanistic'' explanation, although often craved for,
might be rather inadequate/premature \chg[]{here, and will} require better theoretical
understanding \chg[]{of} the demonstrated potential effect of
\chg[]{the heart} anisotropy on cardiac re-entry dynamics, for it is not a particular feature,
or a sequence of anisotropy features, but \chg[mechanistic]{rather the
whole complex of the shape, anisotropy, and the exact heart position within the body
surrounding, which affects the re-entry dynamics in a
particular way, and which
seems to have had evolved in order to ensure the fastest
self-termination of cardiac re-entry. If our hypothesis is correct, it
might explain the difficulties with reproducibility of the arrhythmia
in vivo and in an isolated heart.} \chg[novelty]{The most important
novel finding of the paper is
that, 
contrary to what currently seems to be a commonly accepted 
view of the pro-arrhythmic nature of cardiac anisotropy, the point of
view 
based on the mainly theoretical and simplified anatomy models studies, for the
first time ever, and for the first time in a real whole heart DT-MRI
based model,  we have demonstrated that the real life heart anisotropy
might have 
rather an anti-arrhythmic effect, as it facilitates the fastest self-termination of cardiac
re-entry. }

\begin{acknowledgments}
We acknowledge the support of the UK Medical Research Council grant
G1100357 for the human foetal heart DT-MRI data sets.
We also wish to acknowledge the support of the BeatBox software development
project by EPSRC (UK) grants EP/I029664 and EP/P008690/1.
We thank all the developers of the BeatBox HPC Simulation Environment
for Biophysically and Anatomically Realistic Cardiac
Electrophysiology. We are grateful to  Professor V.N.Biktashev for
much appreciated advice and discussion.
\end{acknowledgments}

%

\end{document}